\documentclass[pdflatex,sn-mathphys-num]{sn-jnl}

\usepackage{graphicx}%
\usepackage{multirow}%
\usepackage{amsmath,amssymb,amsfonts}%
\usepackage{amsthm}
\usepackage[title]{appendix}%
\usepackage{xcolor}%
\usepackage{textcomp}%
\usepackage{manyfoot}%
\usepackage{booktabs}%
\usepackage{algorithm}%
\usepackage{algorithmicx}%
\usepackage{algpseudocode}%
\usepackage{listings}%
\usepackage{txfonts}
\usepackage{bm}
\usepackage{tabularx}
\usepackage{geometry}
\begin{document}

\title[Article Title]{Probing intermediate-mass black hole binaries with the
Lunar Gravitational-wave Antenna}


\author*[1]{\fnm{Hanlin} \sur{Song}}

\author[2,3]{\fnm{Han} \sur{Yan}}

\author[2,3]{\fnm{Yacheng} \sur{Kang}}

\author[2,3]{\fnm{Xian} \sur{Chen}}

\author[4]{\fnm{Junjie} \sur{Zhao}}

\author*[3,5]{\fnm{Lijing} \sur{Shao}}\email{lshao@pku.edu.cn}

\affil[1]{\orgdiv{School of Physics}, \orgname{Peking University},
\orgaddress{\city{Beijing}, \postcode{100871}, \country{China}}}

\affil[2]{\orgdiv{Department of Astronomy}, \orgname{School of Physics, Peking
University}, \orgaddress{\city{Beijing}, \postcode{100871}, \country{China}}}

\affil[3]{\orgdiv{Kavli Institute for Astronomy and Astrophysics},
\orgname{Peking University}, \orgaddress{\city{Beijing}, \postcode{100871},
\country{China}}}

\affil[4]{\orgdiv{Institute for Gravitational Wave Astronomy}, \orgname{Henan
Academy of Sciences}, \orgaddress{\city{Zhengzhou}, \postcode{450046},
\state{Henan}, \country{China}}}

\affil[5]{\orgdiv{National Astronomical Observatories}, \orgname{Chinese Academy
of Sciences}, \orgaddress{\city{Beijing}, \postcode{100012},  \country{China}}}


\abstract{New concepts for observing the gravitational waves (GWs) using a
detector on the Moon, such as the Lunar Gravitational-wave Antenna (LGWA), have
gained increasing attention.  By utilizing the Moon as a giant antenna, the LGWA
is expected to detect GWs in the frequency range from 1 millihertz (mHz) to
several hertz,  with optimal sensitivity in the decihertz band. Despite the
debated formation and evolution channel of intermediate-mass black holes (IMBHs)
with masses in the range of $[10^2, 10^5]\ {\rm M_\odot}$, binary systems
containing at least one IMBH are widely believed to generate GWs spanning from
mHz to a few Hz, making them a key scientific target for the LGWA. We explore
the detectability of IMBH binaries with the LGWA in this work. The LGWA is more
sensitive to nearby binaries (i.e. with redshift $z\lesssim0.5$) with the
primary mass $m_1 \in [10^4, 10^5] \ {\rm M_\odot}$, while it prefers distant
binaries (i.e. $z \gtrsim 5$) with $m_1 \in [10^3, 10^4] \ {\rm M_\odot}$.
Considering a signal-to-noise ratio threshold of 10, our results imply that the
LGWA can detect IMBH binaries up to $z \sim \mathcal{O}(10)$.  We further show
that the LGWA can constrain the primary mass with relative errors $\lesssim
0.1\%$ for binaries at $z \lesssim 0.5$.  Furthermore, we show that the IMBH
binaries at $z \lesssim 0.1$ can be used to constrain redshift with relative
errors $\lesssim 10\%$, and those with $m_1 \in [10^4, 10^5] \ {\rm M_\odot}$
can be localized by the LGWA to be within $\mathcal{O} (10)$ $\rm deg^2$.}

\maketitle

\section{Introduction}\label{sec1}

On September 14, 2015, the first gravitational-wave (GW) event was detected by
the ground-based  Laser  Interferometer Gravitational-wave Observatory (LIGO)
\cite{LIGOScientific:2016aoc}.  More recently, several pulsar timing arrays
(PTAs) have reported intriguing evidence of the Hellings-Downs correlation from
GW signals in the nanohertz band, including the North American Nanohertz
Observatory for Gravitational waves (NANOGrav) \cite{NANOGrav:2023hde,
NANOGrav:2023gor}, the European PTA (EPTA) along with the Indian PTA (InPTA)
\cite{EPTA:2023sfo, EPTA:2023akd, EPTA:2023fyk}, the Parkes PTA (PPTA)
\cite{Zic:2023gta, Reardon:2023gzh}, and the Chinese PTA (CPTA)
\cite{Xu:2023wog}.  Additionally, numerous other GW observatories are currently
under investigation. These include next-generation (XG) ground-based detectors
such as the Einstein Telescope (ET) \cite{Punturo:2010zz} and the Cosmic
Explorer (CE) \cite{Reitze:2019iox}, as well as space-borne detectors such as
LISA \cite{LISA:2017pwj}, Taiji \cite{Hu:2017mde}, TianQin
\cite{TianQin:2015yph}, and DECIGO \cite{Kawamura:2011zz}.

Recently, the new Moon-based detectors, such as the Lunar Gravitational-wave
Antenna (LGWA) \cite{LGWA:2020mma,  Ajith:2024mie}, have gained increasing
attention. When GWs pass by the Moon, it will vibrate,  behaving like a giant
antenna. The LGWA aims to deploy an array of inertial sensors in the Moon's
permanently shadowed regions to monitor its response. Given the exceptionally
quiet and thermally stable environment of the permanently shadowed regions on
the Moon, the LGWA is expected to observe GWs in the frequency range from 1
millihertz (mHz) to several hertz (Hz), with the optimal sensitivity in the
decihertz band.  It will bridge the gap between space-borne detectors with their
optimal sensitivity at the mHz band like LISA, Taiji, and TianQin, and
ground-based detectors like CE and ET with their optimal sensitivity at the
audio band. Meanwhile, parallel design of a Moon-based detector was proposed by
\citet{Li:2023plm} to make use of Chinese lunar exploration project, and it
becomes a good supplement to other projects.

On the other hand, the intermediate-mass black holes (IMBHs) with masses between
$10^2 \ {\rm M_{\odot}}$ and $10^5 \ {\rm M_{\odot}}$ are theoretically believed
to play a crucial role in understanding the evolution of black holes and
dynamics of stellar systems \cite{Greene:2019vlv}. The existence of IMBHs is
indicated by electromagnetic observations in globular clusters, ultraluminous
X-ray sources, and dwarf galaxies \cite{Mezcua:2017npy}. With equipment of
ground-based, space-borne, and Moon-based GW detectors, the binary systems of
IMBHs can be observed through GW experiments. Due to the wide mass range of
IMBHs, the inspiral, merger, and ringdown phases of their binary coalescence can
be observed across the mHz to audio bands \cite{Maggiore:2007ulw}. Recent
studies have explored the detectability of IMBHs across different frequency
bands:  in the mHz range with LISA \cite{Will:2004fj, Arca-Sedda:2020lso,
Strokov:2023kmo, Liu:2023zea}, in the decihertz range with DECIGO
\cite{Sedda:2019uro}, and in the audio band with current \cite{Graff:2015bba,
Veitch:2015ela, Han:2017evx, Liu:2023zea} and XG \cite{Huerta:2010tp,
Arca-Sedda:2020lso, Reali:2024hqf} ground-based GW detectors. Given that the
Moon-based GW detectors have the optimal sensitivity in the decihertz band, the
IMBHs also become a potentially key science target of those projects
\cite{LGWA:2020mma, Ajith:2024mie, Li:2023plm}.  

In this work, we focus on the detectability of IMBH binaries with quasi-circular
orbits and aligned spins with the LGWA. Following the approach by
\citet{Reali:2024hqf}, we perform a parameter scan with straightforward priors
to generate the masses and redshifts of IMBH binaries. We then calculate the
distribution of signal-to-noise ratios (SNRs) with different masses and
redshifts for IMBH binaries. Using the Fisher information matrix (FIM) method,
we calculate the relative parameter inference errors for some key parameters,
such as the primary mass and redshift. We also calculate the localization
capability of LGWA on IMBH binaries. 

This paper is arranged as follows. In Sec.~\ref{settings_methods}, we introduce
the methodology and settings adopted in this work, including the parameter
priors of IMBH binaries, the GW waveform template, the LGWA configuration, and a
brief introduction of the FIM method. In Sec.~\ref{results}, we present our
results on the detectability of IMBH binaries by the LGWA. Finally, we conclude
in Sec.~\ref{summary}. Throughout the paper, we adopt natural units $G=c=1$.

\section{Settings and Methods}
\label{settings_methods}

In this section, we outline the settings and methods adopted in this work. In
Sec.~\ref{population}, we describe the parameter priors of IMBH binaries. In
Sec.~\ref{waveform}, we introduce the waveform model and the LGWA
configurations. In Sec.~\ref{fisher}, we briefly discuss the FIM method for
parameter estimation. 

\subsection{Priors for the population of IMBH binaries}
\label{population}

For the IMBH binary systems with aligned spins and quasi-circular orbits, the GW
signals from them are described with 11 free parameters,
\begin{align}
	\bm{\theta} = \Big\{ m_1,\ m_2,\ \chi_{1z},
	\ \chi_{2z},\  D_L, \ \alpha, 
	 \ \delta , \ \psi , \ \iota, \ \phi_c,
	\ t_c  \Big\},
\end{align}
where $m_1$ and $m_2$ are source-frame binary masses of the two components,
$\chi_{1z}$ and $\chi_{2z}$ are their dimensionless spin components which
parallel with the orbital angular momentum, $D_L$ is the luminosity distance of
the source, $\alpha$ and $\delta$ are the right ascension and declination angles
respectively, $\iota$ and $\psi$ are the inclination angle the polarization
angle respectively, and $\phi_c$ and $t_c$ are the coalescence phase and time
respectively. 

Given the debated astrophysical star formation rate and merger rate for IMBHs,
we follow \citet{Reali:2024hqf} to adopt sample parameter priors of IMBHs which
span the parameter space.  As shown in Table~\ref{prior_table}, the source-frame
primary mass is sampled logarithmically uniform from $m_{1} \in [10^2, 10^5]$
$\rm M_{\odot}$, while the secondary mass is sampled logarithmically uniform
from $m_{2} \in [10 \ \rm  M_{\odot}, m_1]$. Meanwhile, the mass ratio
$q=m_1/m_2 \in [1, 10]$ is further imposed in this work to guarantee the
accuracy of the waveform model \cite{Reali:2024hqf, Garcia-Quiros:2020qpx}. We
consider IMBH binaries which are fixed at six representative redshifts, $z \in
\big\{ 0.05, 0.1, 0.5, 1, 5, 10 \big\}$, to cover both nearby and distant cases.
The luminosity distance $D_L$ is then obtained using the $\Lambda$CDM cosmology
model. The four angles $\alpha$, $\cos{\delta}$,  $\psi$, and $\cos{\iota}$,
follow the uniform distributions  $\mathcal{U}[0, 2\pi)$, $\mathcal{U}[-1,1]$,
$\mathcal{U}[0, \pi]$, and $\mathcal{U}[-1,1]$, respectively. The priors for
the coalescence phase, coalescence time, and two aligned spins are fixed to zero
for all cases. 

\begin{table}[bp]
\renewcommand{\arraystretch}{1.3}
  \centering
  \caption{Parameter priors of IMBH binaries.}
    \begin{tabularx}{\textwidth}{XX}
    \toprule
    Parameter & Priors   \\
    \toprule
    $m_{1}$ & [$10^2$, $10^5$] $\rm M_{\odot}$ in log-uniform \\
    $m_{2}$ & $[10 \ {\rm M_{\odot}}, m_1]$ \  in log-uniform \\
    $z$ & \big\{0.05, 0.1, 0.5, 1, 5, 10 \big\} \\
    $\alpha$ & $\mathcal{U}[0, 2\pi]$ \\
    $\cos{\delta}$ & $\mathcal{U}[-1,1]$ \\
    $\psi$ & $\mathcal{U}[0, \pi]$ \\
    $\cos{\iota}$ &  $\mathcal{U}[-1,1]$ \\
    $\phi_c$ & 0 \\
    $t_c$ & 0 \\
     $\chi_{1z,2z}$ & 0 \\
    \bottomrule
    \end{tabularx}
  \label{prior_table}
\end{table}

\subsection{Waveform model and detectors}
\label{waveform}

The detected GW strain is described as \cite{whelan2013geometry},
\begin{equation}
    h = h_{+} \stackrel{\leftrightarrow}{e}_{+} : \stackrel{\leftrightarrow}{d}
    + h_{\times} \stackrel{\leftrightarrow}{e}_{\times} :
    \stackrel{\leftrightarrow}{d},
\end{equation}
where $h_+$ and $h_\times$ are the `+' and `$\times$' polarization components of
the GW respectively, $\stackrel{\leftrightarrow}{e}_{+}$ and
$\stackrel{\leftrightarrow}{e}_{\times}$ are the corresponding polarization
tensors respectively, and $\stackrel{\leftrightarrow}{d}$ is the response tensor
of the detector. 

Unlike current ground-based ``L'' shape laser interferometer type detectors like
LIGO, Virgo, and KAGRA, the detection principle of LGWA has significant
differences \cite{LGWA:2020mma, Branchesi:2023sjl, Ajith:2024mie}. When a GW
passes the Moon, its surface displacements can be measured by several
seismometers of the LGWA. With the long-wavelength approximation, the response
tensor of a seismometer is written as \cite{Dupletsa:2022scg},  
\begin{equation}
    \stackrel{\leftrightarrow}{d}=\vec{e}_n\otimes\vec{e}_1,
    \label{eq:resptensor}
\end{equation}
where $\vec{e}_n$ is the direction of the surface normal vector at the
seismometer, and $\vec{e}_1$ is the direction of displacement measured by the
seismometer.  For a more realistic situation, the complete result of the
response tensor is somewhat more complicated, including different contributions
of radial and horizontal vibrations \cite{Yan:2024sgwb}. Thus,
Eq.~(\ref{eq:resptensor}) can be regarded as a simplified case in which the
radial response function is just two times of the horizontal response function,
as shown in Eq.~(7) in \citet{Yan:2024sgwb}.

We use the \texttt{IMRPhenomXHM} waveform template \cite{Garcia-Quiros:2020qpx}
to generate  $h_+$ and $h_\times$. Additionally, we employ the \texttt{GWFish}
package \cite{Dupletsa:2022scg} to calculate the GW strain projected onto the
seismometer. Note that  \texttt{GWFish} also takes into considerations the
orbital motion of the Moon around the Earth, as well as the motion of the
Earth-Moon system around the Sun during the detection period. We focus on the GW
frequency band ranging from $10^{-3}$ Hz to 4 Hz. Additionally, 20,000 points
are evenly spaced on a logarithmic scale for each signal. Furthermore, we
utilize the default \texttt{LGWA} detector in \texttt{GWFish}, which consists of
an array of four stations deployed in the Moon's permanently shadowed regions.
Each station is equipped with two horizontal Lunar inertial GW sensors, which
measure the two orthogonal surface displacements \cite{LGWA:2020mma,
Branchesi:2023sjl, Ajith:2024mie}. The mission duration of the LGWA is expected
to be 10 years. Thus, GW signals with time durations longer than 10 years are
truncated at a low-frequency end via \cite{Dupletsa:2022scg},
\begin{equation}
t(f) = t_c - \frac{5}{256\mathcal{M}_c^{5/3}}( \pi f)^{-8/3}, 
\end{equation} 
where $\mathcal{M}_c$ is the chirp mass.

\subsection{Fisher information matrix}
\label{fisher}

Under the linear-signal approximation and assuming  Gaussian and stationary
noise, the posterior distribution of GW parameters becomes  \cite{Finn:1992wt,
Borhanian:2020ypi},
\begin{equation}\label{eq12}
	p(\boldsymbol{\theta})\sim
	\mathrm{e}^{-\frac12\Gamma_{ij}\Delta\theta_{i}\Delta\theta_{j}},
\end{equation}
where  $\Gamma_{ij}$ is the FIM and can be calculated as,
\begin{equation}
	\Gamma_{ij}\equiv \Big\langle\partial_{\theta_{i}}
	h(\boldsymbol{\theta};f),\partial_{\theta_{j}} h(\boldsymbol{\theta};f)
	\Big\rangle.
\end{equation}
Note that the inner product for two quantities  $A(\boldsymbol{\theta};f)$ and
$B(\boldsymbol{\theta};f)$ is defined as,
\begin{equation}
	\langle A,
	B\rangle=2\int_{0}^{\infty}\mathrm{d}f\frac{A(\boldsymbol{\theta};f)
	B^{*}(\boldsymbol{\theta};f)+A^{*}(\boldsymbol{\theta};f)
	B(\boldsymbol{\theta};f)}{S_{n}(f)},
\end{equation}
where ${S_{n}(f)}$ is the one-sided power spectrum density (PSD) of the
detector.  The matched-filtering SNR for a GW event is calculated as,
\begin{align}
    \label{SNR}
    {\rm  SNR}= \sqrt{\langle h ,h \rangle}.
\end{align}

We also use the \texttt{GWFish} package for FIM calculations. 

\begin{figure*}[t] 
	\centering 
    \includegraphics[width=9cm]{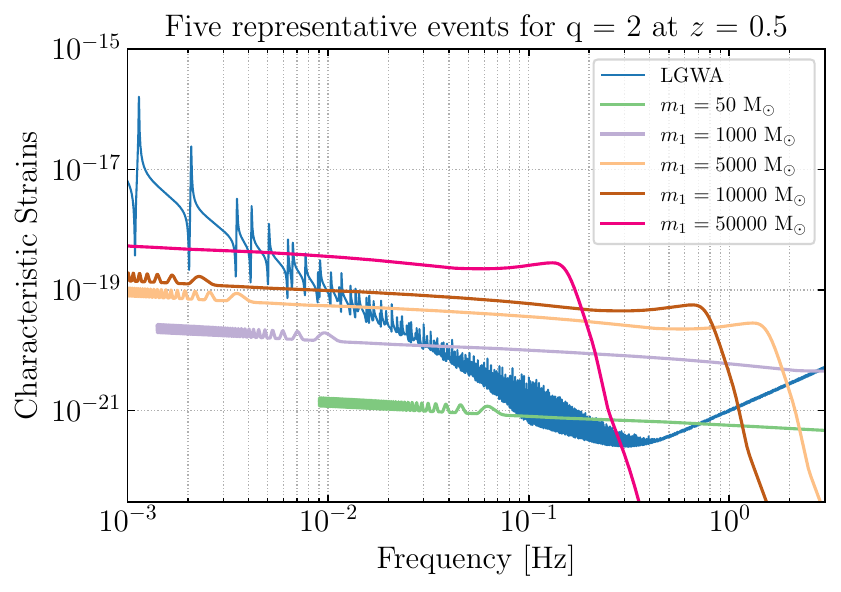}
    \hspace{1em}
    \centering
    \includegraphics[width=7cm]{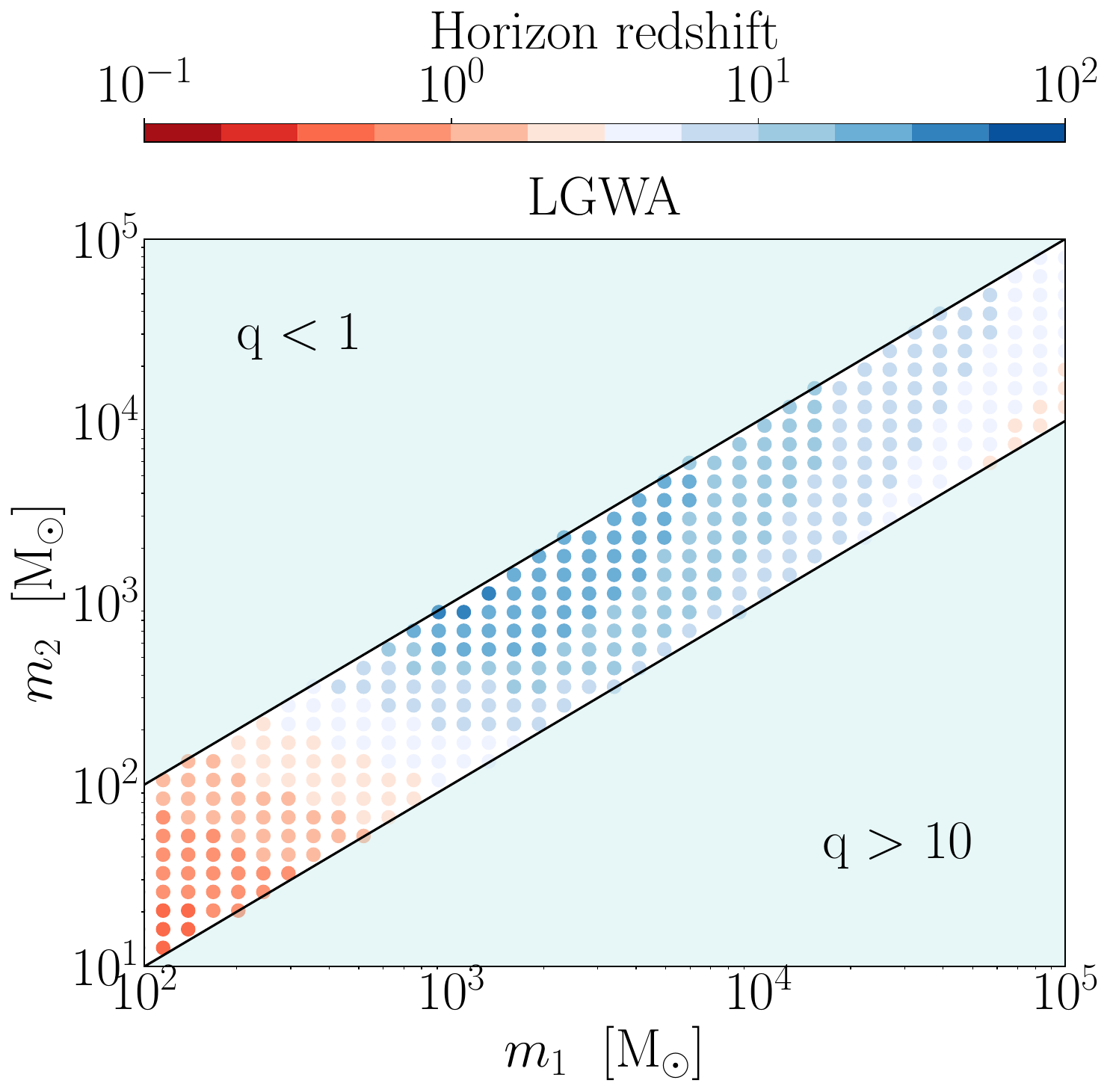} 
    \caption{The left panel shows the characteristic strain $h_c$ for five
    representative events with different primary masses. The characteristic
    noise $h_n$ of the LGWA is presented in blue line. All events are fixed at
    $z=0.5$ with a mass ratio $q=2$. The right panel shows the horizon redshift
    with a detection threshold  ${\rm SNR}=10$. The region with mass ratio $q<1$
    is shaded for the requirement of $m_1 \geq m_2$, while the one with $q>10$
    is shaded for the applicability of the waveform model. For both panels, the
    four free angles in Table~\ref{prior_table} are fixed to  $\alpha = \pi/4$,
    $\cos{\delta}=1/2$,  $\psi=\pi/4$, and $\cos{\iota}=1$.} 
    \label{character_strains}
\end{figure*}

\begin{figure*}[t] 
	\centering 
	\includegraphics[width=1\textwidth]{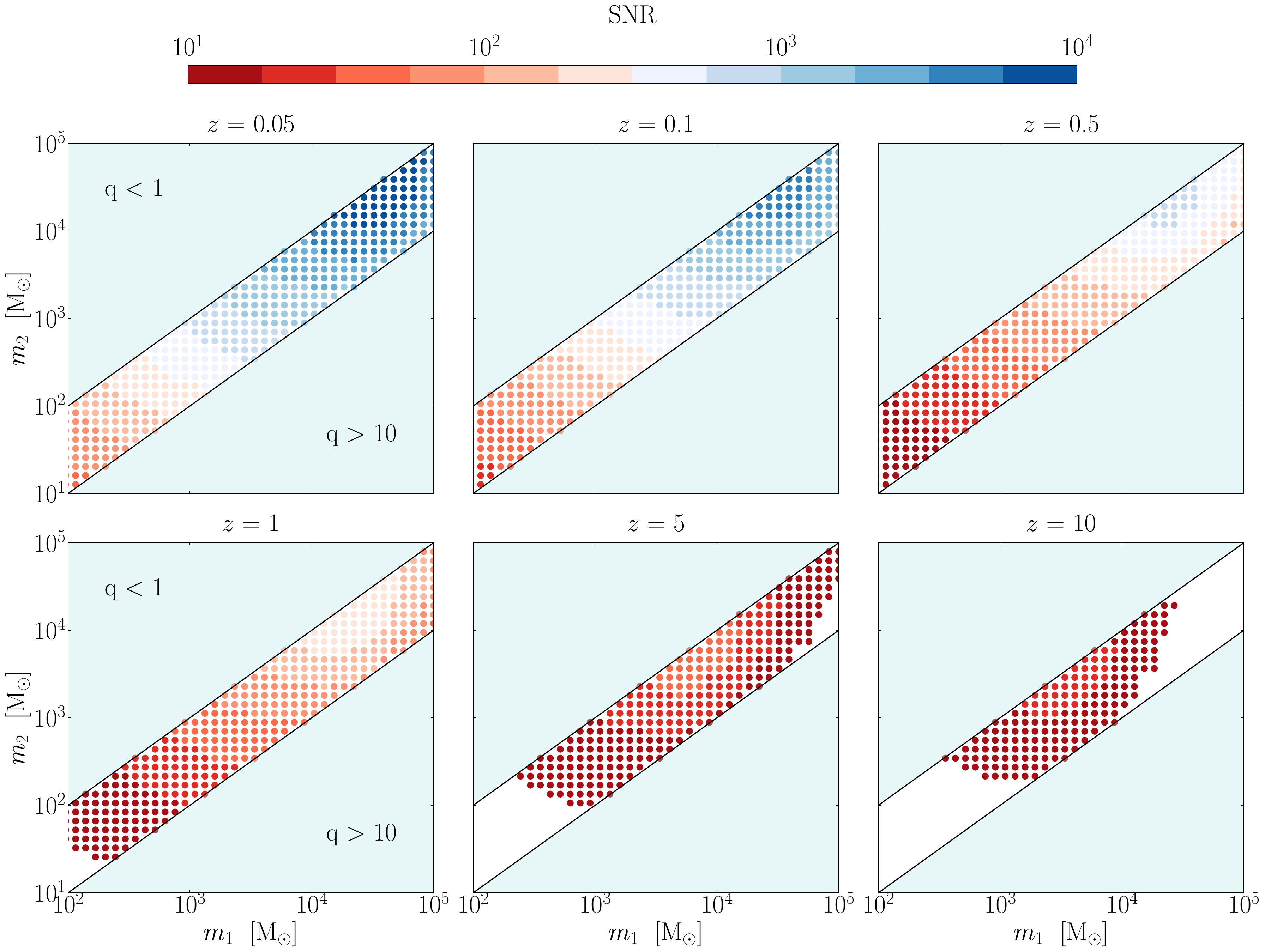}
	\caption{Each subfigure shows the angle-averaged SNR for binaries fixed at
	different redshifts. For each point, we take an average of 1000 events which
	are sampled with priors shown in Table~\ref{prior_table}. The blank region
	with no points in the lower subfigures represents the binary systems with
	angle-averaged SNR $<10$.} 
	\label{Fig_SNR} 
\end{figure*}
\begin{figure*}[t] 
	\centering 
	\includegraphics[width=0.6\textwidth]{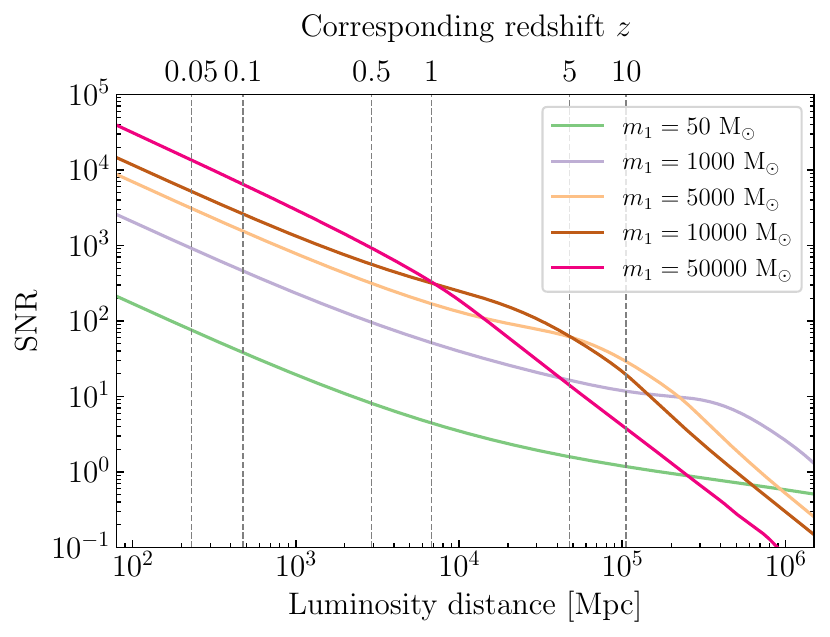} 
	\caption{The evolution of SNR with respect to luminosity distance. These
	five events have same parameters as in Fig.~\ref{character_strains} with
	corresponding color.} 
	\label{Fig_SNR_with_DL} 
\end{figure*}

\section{Results}
\label{results}

With the definition of SNR shown in Eq.~\eqref{SNR}, the characteristic strains for a GW event $h_{c}$ and the characteristic noise $h_n$ can be defined as \cite{Moore:2014lga},
\begin{equation}
\begin{aligned}
    h_{c} &= 2f|h|, \\
    h_{n} &= \sqrt{f S_n}.
\end{aligned}
\end{equation}

In the left panel of Fig.~\ref{character_strains}, we plot the characteristic
strains for five representative events of IMBH binaries and the characteristic
sensitivity of LGWA. These events are put at $z=0.5$ with five different primary
masses $m_1 \in \big\{50, 1000, 5000, 10000, 50000 \big\} \ {\rm M_\odot}$. 
Additionally, the mass ratio is fixed to $q=2$, and the four free angle
parameters are fixed to $\alpha = \pi/4$, $\cos{\delta}=1/2$,  $\psi=\pi/4$, and
$\cos{\iota}=1$. As we can see, there exists a low-frequency cut off for lighter
binaries, which is determined by the assumed maximum  observation time of 10
years.  The sinusoidal fluctuation at the low-frequency band actually arises
from the orbital motion of the Moon around our Earth, as well as the motion of
the Earth-Moon system around the Sun.  With the primary mass increasing, the
amplitude of GW signal increases, while the chirp frequency decreases.  In the
right panel of Fig.~\ref{character_strains}, we plot the horizon redshift for
IMBH binaries with mass ratio between 1 and 10. The four angles are also fixed
as in the left panel. The detection threshold is chosen as $\rm{SNR} = 10$. We
can clearly see that the binaries with primary mass $m_1 \in [10^3, 10^4] \ {\rm
M_\odot}$ have the best performance, which can be detected with a horizon
redshift of $z \sim \mathcal{O}(10)$. These results are consistent with 
\citet{Ajith:2024mie}.

Then we calculate the angle-averaged SNR for IMBHs at $z \in \big\{0.05, 0.1,
0.5, 1, 5, 10 \big\}$. The results are shown in Fig.~\ref{Fig_SNR}. For each
point in the figure, we generate 1000 events with the corresponding $m_1$,
$m_2$, and $z$ values, while the four free angle parameters $\alpha$,
$\cos{\delta}$,  $\psi$, and $\cos{\iota}$ are randomly drawn from the uniform
priors in Table~\ref{prior_table}. The angle-averaged SNR is obtained by taking
the average of the results from 1000 events. We find that all events can be
detected at $z\lesssim0.5$. However, at redshift $z\gtrsim1$,  the binaries with
primary mass $m_1 \in [10^2, 10^3] \ {\rm M_\odot}$ and $m_1 \in [10^4, 10^5] \
{\rm M_\odot}$ show less detection efficiency.  The angle-averaged SNRs fall
below 10 for binaries at $z=10$. It is interesting to note that for nearby
binaries (i.e. $z \lesssim 0.5$), the binaries with $m_1 \in [10^4, 10^5] \ {\rm
M_\odot}$ show better performance. For example, at $z=0.05$ or $z=0.1$, binaries
with $m_1 \in [10^4, 10^5] \ {\rm M_\odot}$ can be detected with angle-averaged
SNR $>\mathcal{O}(10^3)$, while for distant systems  (i.e. $z\gtrsim5$),
binaries with $m_1 \sim [10^3, 10^4]\ {\rm M_\odot}$ show  better performance. 
This phenomenon arises due to the sensitivity profile of LGWA in the decihertz
band. In Fig.~\ref{Fig_SNR_with_DL}, we plot the evolution of SNR with
luminosity distance for each of the above five representative events in
Fig.~\ref{character_strains}. The pink line denotes the binary system with $m_1
= 50000 \ {\rm M_\odot}$, while the light orange line denotes the system with
$m_1 = 5000 \ {\rm M_\odot}$. As the luminosity distance increases, the LGWA
becomes sensitive first to heavier systems and then to lighter systems. The
turning point happens around $D_L \sim 10^4$ Mpc, which corresponds to $z \sim
1.5$ in the $\Lambda$CDM cosmology model. The SNR evolution for systems with
other primary mass are also consistent with the results shown in
Fig.~\ref{Fig_SNR}. It is worth noting that the variation in detection
sensitivity with source-frame mass and redshift can be mitigated by converting
$m_1$ in Fig.~\ref{Fig_SNR} to detector-frame mass. For instance, in the cases
of $z = 5$ and $z = 10$, the detector-frame mass will be magnified by
approximately $\sim \mathcal{O}(10)$, whereas other cases remain almost
unaffected.

\begin{figure*}[t] 
	\centering 
	\includegraphics[width=1\textwidth]{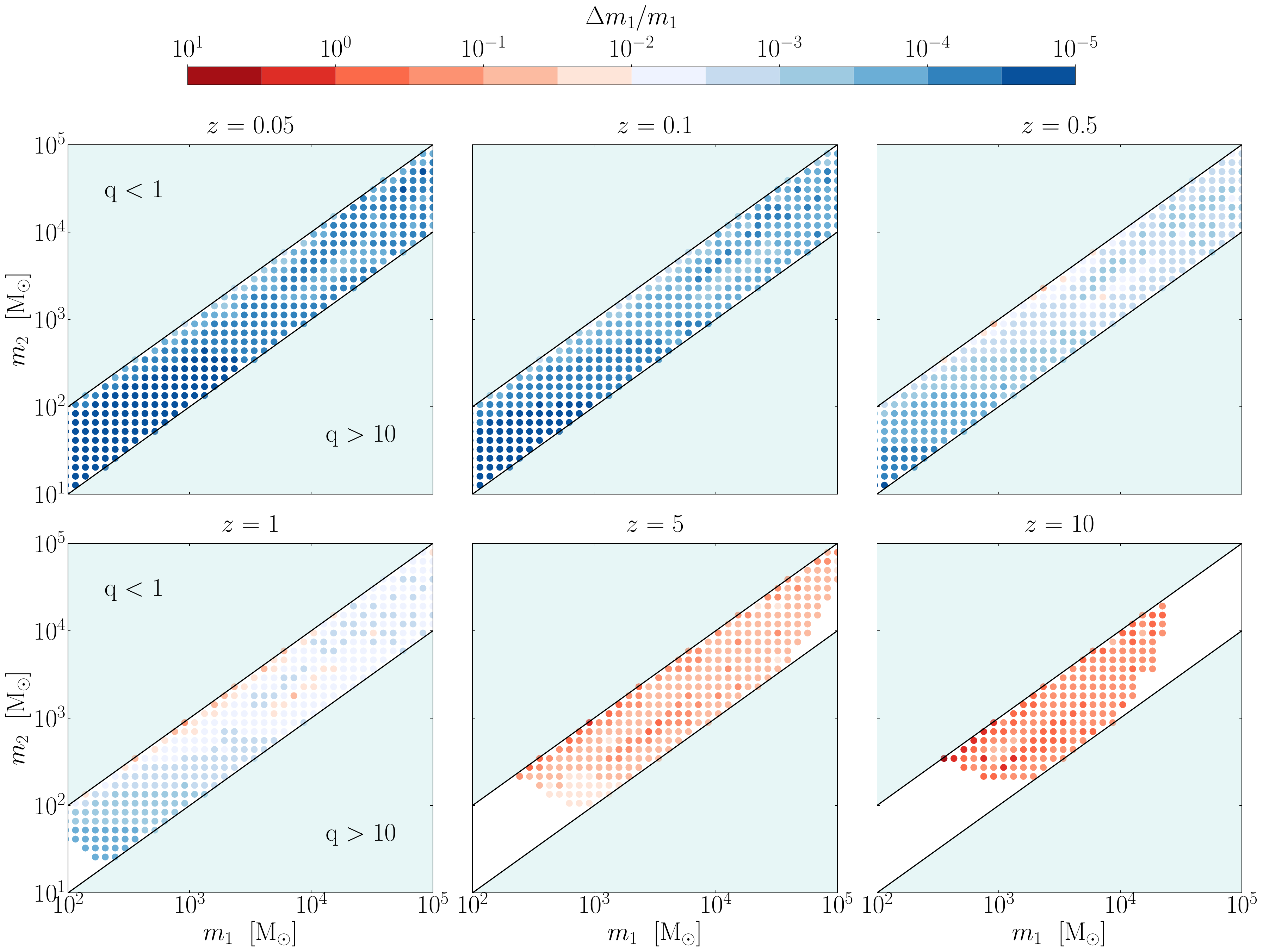} 
	\caption{Same as Fig.~\ref{Fig_SNR}, but for the angle-averaged relative
	error of the primary mass, $\Delta m_1/m_1$.} 
	\label{Fig_errs_m1} 
\end{figure*}
\begin{figure*}[t] 
	\centering 
	\includegraphics[width=1\textwidth]{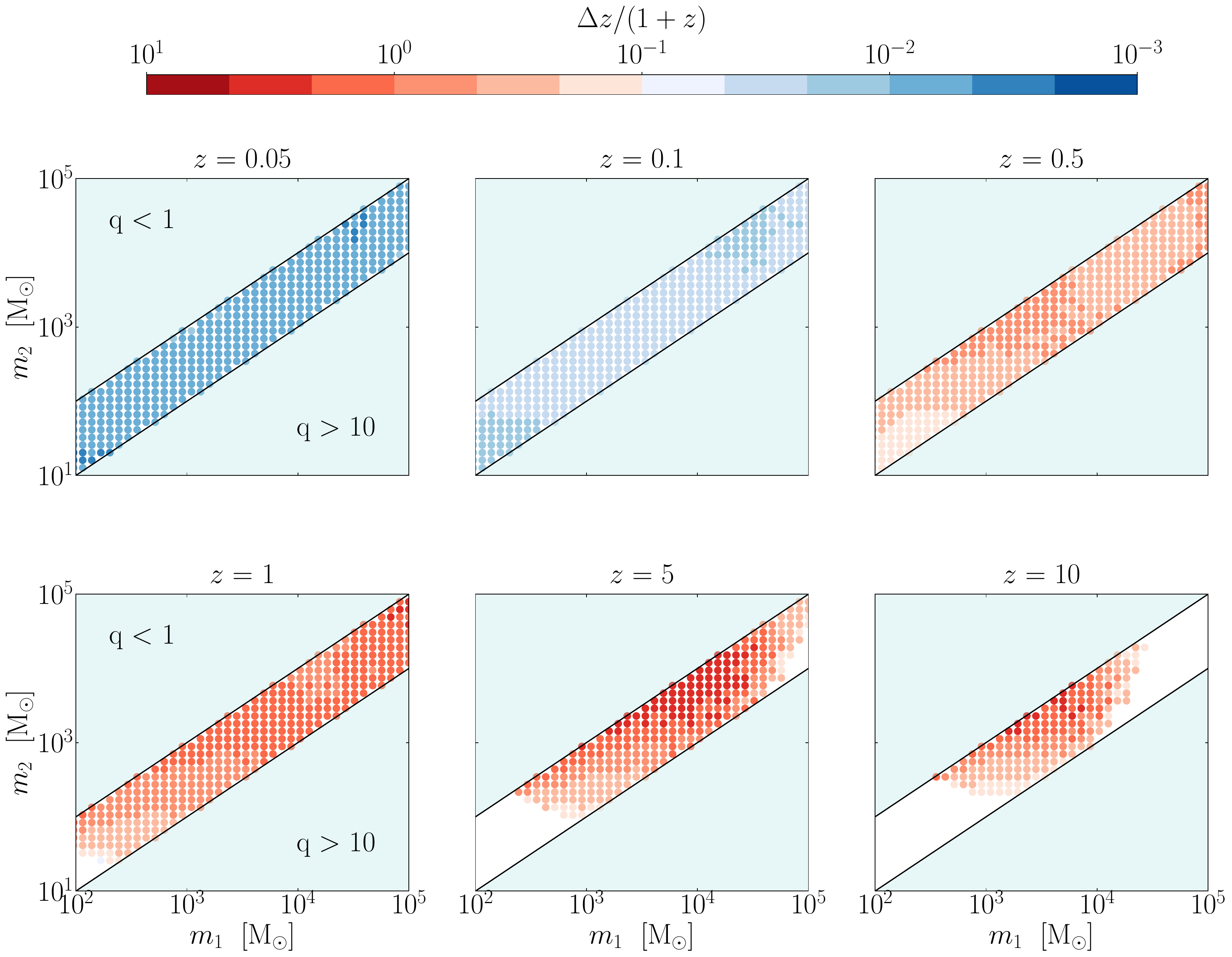} 
	\caption{Same as Fig.~\ref{Fig_SNR}, but for the angle-averaged relative
	error of the redshift,  $\Delta z/(1+z)$.} 
	\label{Fig_errors_redshift} 
\end{figure*}
\begin{figure*}[htbp] 
	\centering 
	\includegraphics[width=0.65\textwidth]{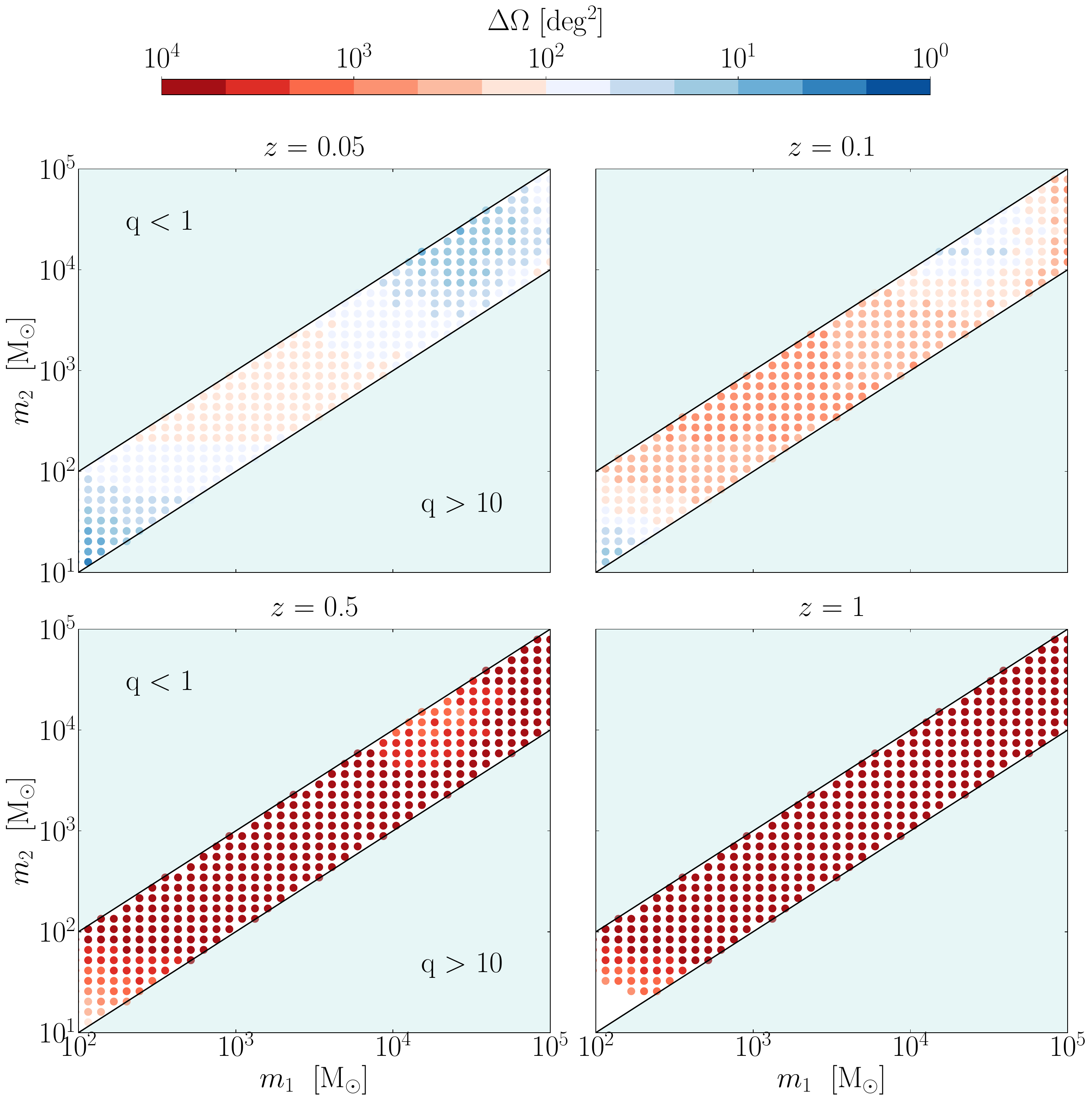} 
	\caption{Same as Fig.~\ref{Fig_SNR}, but for the angle-averaged 90\% sky
	localization errors at different redshifts, $z \in \big\{0.05, 0.1, 0.5, 1
	\big\}$.} 
	\label{Fig_errors_sky_localization} 
\end{figure*}

In the following, we estimate the relative errors for $m_1$, $z$, and the $90\%$
sky localization uncertainty of IMBH binaries with the LGWA. In
Fig.~\ref{Fig_errs_m1}, we show the angle-averaged relative errors of the
primary mass $\Delta m_1/m_1$  at different redshifts. For nearby binaries (i.e.
$z\lesssim0.5$), the primary mass can be measured with accuracy better than
$0.1\%$. For binaries at $z=1$, the primary mass can be measured with
uncertainty less than $1\%$ for most cases. However, for distant binaries (i.e.
$z\gtrsim5$), the primary mass can only be measured with lower accuracy,
typically worse than $10\%$ across most detectable regions. 
Figure~\ref{Fig_errors_redshift} shows results for angle-averaged relative error
of the redshift,  $\Delta z/(1+z)$.  For binaries at $z\lesssim0.1$, the
redshift can be constrained with relative errors $\lesssim10\%$. For binaries at
$z=0.05$, the redshift can be constrained better than $1\%$. However, for
binaries at $z\gtrsim0.5$, the redshift is constrained worse than $10\%$ for all
detectable regions.  In Fig.~\ref{Fig_errors_sky_localization}, we plot the
angle-averaged 90\% sky localization uncertainty \cite{Li:2021mbo,
Dupletsa:2022scg} for IMBH binaries at different redshifts. For binaries at
$z\lesssim0.1$, those with $m_1 \in [10^4, 10^5] \ {\rm M_\odot}$ can be
localized within $\mathcal{O}(10) \ \rm{deg}^2$, while binaries with $m_1 \in
[10^3, 10^4] \ {\rm M_\odot}$ can be localized within around $\mathcal{O}(10^2)
\ \rm{deg}^2$. However, for binaries at $z>0.5$, all events show poor
localization precision.  

\section{Summary}
\label{summary}

\begin{figure*}[htpb] 
	\centering 
    \includegraphics[width=9cm]{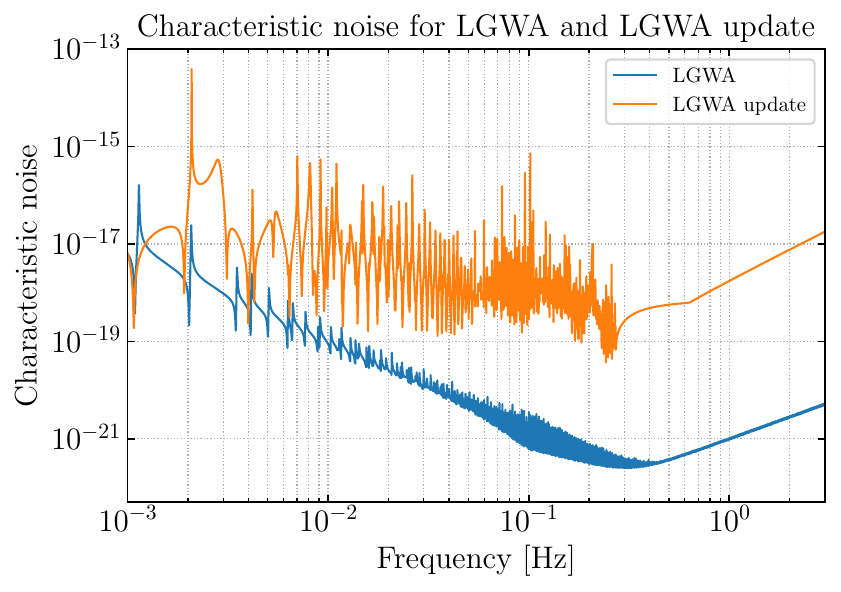}
	\hspace{1.5em}
    \centering
    \includegraphics[width=7cm]{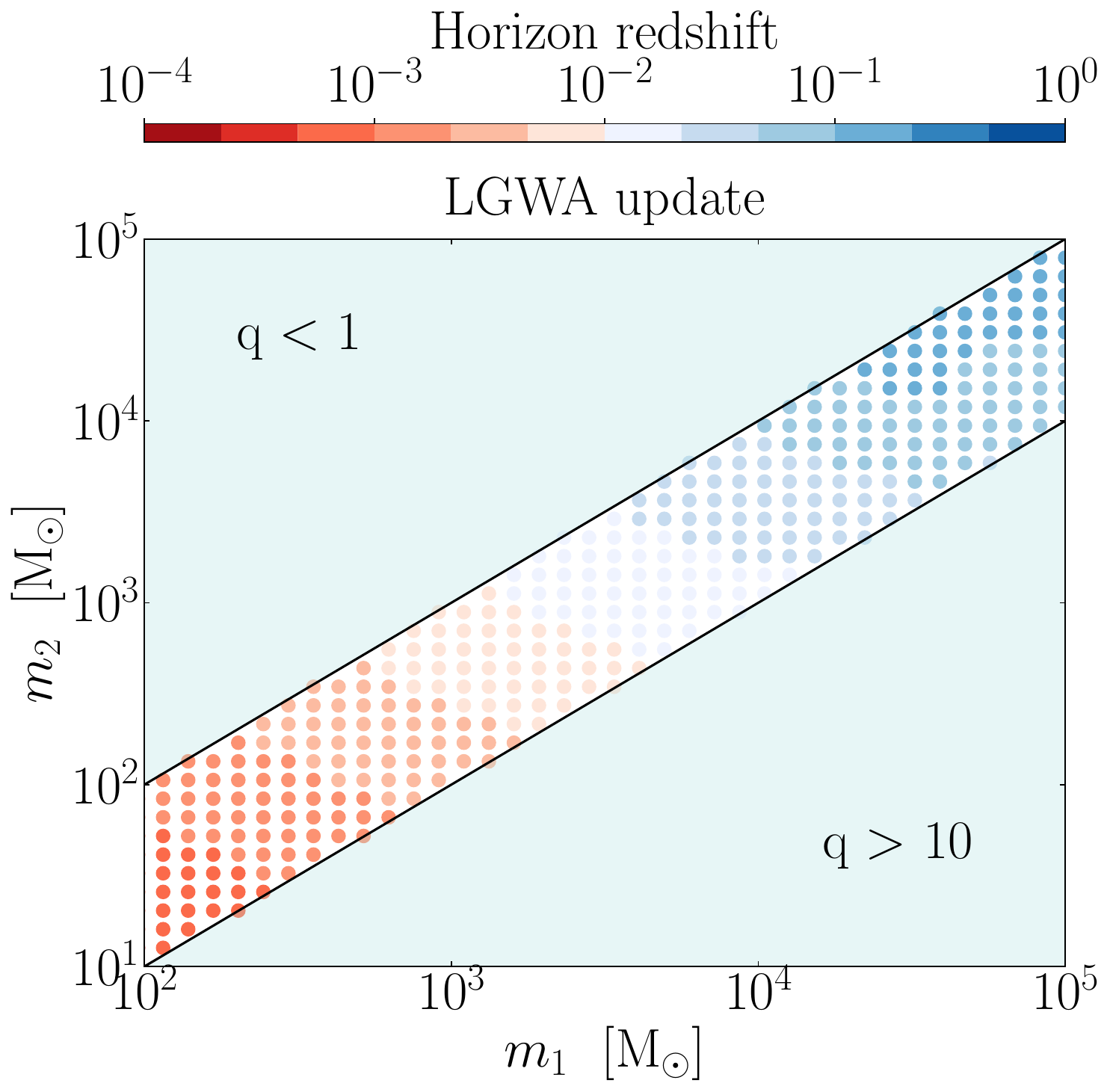}
    \caption{The left panel shows the characteristic noise $h_n$ for LGWA and
    LGWA update. The right panel shows the horizon redshift same to
    Fig.~\ref{character_strains} but here for LGWA update.} 
    \label{character_PSD_strains}
\end{figure*}

We explored the detectability of IMBH binaries with the LGWA. Due to its unique
shape of the sensitivity curve at decihertz band, the LGWA is more sensitive to
distant binaries (i.e. $z\gtrsim5$) with $m_1 \in [10^3, 10^4] \ {\rm M_\odot}$,
while preferring nearby binaries (i.e. $z\lesssim0.5$) with $m_1 \in [10^4,
10^5] \ {\rm M_\odot}$. The primary mass can be measured with accuracy better
than $0.1\%$ for binaries at $z\lesssim0.5$, and the redshift can be constrained
within $10\%$ for binaries at $z\lesssim0.1$. Meanwhile,  binaries with $m_1 \in
[10^4, 10^5] \ {\rm M_\odot}$ can be localized within $\mathcal{O} (10 ) \ {\rm
deg^2}$ at $z \lesssim 0.1$. As the LGWA fills the unexplored GW frequency band
between space-borne detectors (i.e., LISA, Taiji, and TianQin) and ground-based
detector (i.e., CE and ET), it shows the unique advantage for detecting IMBH
binaries.  The observations of the inspiral and merger phases of IMBH binaries
can offer us more insights into the formation and evolution of black holes,
especially bridging the gap between stellar-mass and supermassive black holes
\cite{Sesana:2010wy}.  Additionally, it will help us to disentangle the effects
between the accretion history and the merger dynamics of black holes
\cite{Sesana:2007vr, Klein:2015hvg}. 

Several recent studies \cite{Yan:2024jio,Belgacem:2024eft,Majstorovic:2024arXiv}
showed that the PSD of the LGWA should be updated when carefully considering the
lunar response to GWs. We show the updated PSD with the same approach as in
\citet{Yan:2024jio} in the left panel of Fig.~\ref{character_PSD_strains}, but
modify it a little bit using a new sensitivity curve of the LGWA seismometer (we
choose the black dotted line as the sensitivity of a single detector, from
Fig.~2(b) in \citet{Ajith:2024mie}). The updated PSD matches marginally well
with the original one in the frequency region lower than 10 mHz but becomes two
orders of magnitude worse around decihertz. This variation is primarily because
that \citet{Yan:2024jio} used the Dyson-type force density, rather than the
tidal force density, to calculate the lunar response to GWs.  In the right panel
of Fig.~\ref{character_PSD_strains}, we show horizon redshift of IMBH binaries
for the updated LGWA PSD. Because of the change of sensitivity in the decihertz
band, the IMBH binaries now can only be detected up to $z \sim \mathcal{O} (1)$.
Finally, more details, including a more precise calculation of the lunar
response with the near-surface fine structure of the Moon, need to be considered
for future exploration of the science cases with the LGWA.

\bmhead{Acknowledgements} 

We thank Zhenwei Lyu for helpful discussion and comments.  This work was
supported by the Beijing Natural Science Foundation (1242018), the National SKA
Program of China (2020SKA0120300), the Max Planck Partner Group Program funded
by the Max Planck Society, the Fundamental Research Funds for the Central
Universities and the High-performance Computing Platform of Peking University.
J.Z.\ is supported by the National Natural Science Foundation of China (No.
12405052) and the Startup Research Fund of Henan Academy of Sciences (No.
241841220). Y.K.\ is supported by the China Scholarship Council (CSC).
X.C.\ is supported by the Natural Science Foundation of China (No. 12473037).

\bmhead{Author contributions}
H.S.\ contributed to the design, calculation and writing of the initial draft.
Y.H.\ contributed to the sensitivity analysis of the LGWA, and Y.H., Y.K., and
J.Z.\ contributed to discussions and revision of the work.   X.C.\ contributed
to the discussion of the results.  L.S.\ initiated the idea, revised the
manuscript, and provided supervision and coordination of the whole project.  

\bibliography{lunar_ref.bib}


\begin{thebibliography}{46}
\ifx \bisbn   \undefined \def \bisbn  #1{ISBN #1}\fi
\ifx \binits  \undefined \def \binits#1{#1}\fi
\ifx \bauthor  \undefined \def \bauthor#1{#1}\fi
\ifx \batitle  \undefined \def \batitle#1{#1}\fi
\ifx \bjtitle  \undefined \def \bjtitle#1{#1}\fi
\ifx \bvolume  \undefined \def \bvolume#1{\textbf{#1}}\fi
\ifx \byear  \undefined \def \byear#1{#1}\fi
\ifx \bissue  \undefined \def \bissue#1{#1}\fi
\ifx \bfpage  \undefined \def \bfpage#1{#1}\fi
\ifx \blpage  \undefined \def \blpage #1{#1}\fi
\ifx \burl  \undefined \def \burl#1{\textsf{#1}}\fi
\ifx \doiurl  \undefined \def \doiurl#1{\url{https://doi.org/#1}}\fi
\ifx \betal  \undefined \def \betal{\textit{et al.}}\fi
\ifx \binstitute  \undefined \def \binstitute#1{#1}\fi
\ifx \binstitutionaled  \undefined \def \binstitutionaled#1{#1}\fi
\ifx \bctitle  \undefined \def \bctitle#1{#1}\fi
\ifx \beditor  \undefined \def \beditor#1{#1}\fi
\ifx \bpublisher  \undefined \def \bpublisher#1{#1}\fi
\ifx \bbtitle  \undefined \def \bbtitle#1{#1}\fi
\ifx \bedition  \undefined \def \bedition#1{#1}\fi
\ifx \bseriesno  \undefined \def \bseriesno#1{#1}\fi
\ifx \blocation  \undefined \def \blocation#1{#1}\fi
\ifx \bsertitle  \undefined \def \bsertitle#1{#1}\fi
\ifx \bsnm \undefined \def \bsnm#1{#1}\fi
\ifx \bsuffix \undefined \def \bsuffix#1{#1}\fi
\ifx \bparticle \undefined \def \bparticle#1{#1}\fi
\ifx \barticle \undefined \def \barticle#1{#1}\fi
\bibcommenthead
\ifx \bconfdate \undefined \def \bconfdate #1{#1}\fi
\ifx \botherref \undefined \def \botherref #1{#1}\fi
\ifx \url \undefined \def \url#1{\textsf{#1}}\fi
\ifx \bchapter \undefined \def \bchapter#1{#1}\fi
\ifx \bbook \undefined \def \bbook#1{#1}\fi
\ifx \bcomment \undefined \def \bcomment#1{#1}\fi
\ifx \oauthor \undefined \def \oauthor#1{#1}\fi
\ifx \citeauthoryear \undefined \def \citeauthoryear#1{#1}\fi
\ifx \endbibitem  \undefined \def \endbibitem {}\fi
\ifx \bconflocation  \undefined \def \bconflocation#1{#1}\fi
\ifx \arxivurl  \undefined \def \arxivurl#1{\textsf{#1}}\fi
\csname PreBibitemsHook\endcsname

\bibitem[\protect\citeauthoryear{Abbott et~al.}{2016}]{LIGOScientific:2016aoc}
\begin{barticle}
\bauthor{\bsnm{Abbott}, \binits{B.P.}}, \betal:
\batitle{{Observation of Gravitational Waves from a Binary Black Hole Merger}}.
\bjtitle{Phys. Rev. Lett.}
\bvolume{116}(\bissue{6}),
\bfpage{061102}
(\byear{2016})
\doiurl{10.1103/PhysRevLett.116.061102}
{\href{https://arxiv.org/abs/1602.03837}{{arXiv:1602.03837}}}
{[gr-qc]}
\end{barticle}
\endbibitem

\bibitem[\protect\citeauthoryear{Agazie et~al.}{2023a}]{NANOGrav:2023hde}
\begin{barticle}
\bauthor{\bsnm{Agazie}, \binits{G.}}, \betal:
\batitle{{The NANOGrav 15 yr Data Set: Observations and Timing of 68
  Millisecond Pulsars}}.
\bjtitle{Astrophys. J. Lett.}
\bvolume{951}(\bissue{1}),
\bfpage{9}
(\byear{2023})
\doiurl{10.3847/2041-8213/acda9a}
{\href{https://arxiv.org/abs/2306.16217}{{arXiv:2306.16217}}}
{[astro-ph.HE]}
\end{barticle}
\endbibitem

\bibitem[\protect\citeauthoryear{Agazie et~al.}{2023b}]{NANOGrav:2023gor}
\begin{barticle}
\bauthor{\bsnm{Agazie}, \binits{G.}}, \betal:
\batitle{{The NANOGrav 15 yr Data Set: Evidence for a Gravitational-wave
  Background}}.
\bjtitle{Astrophys. J. Lett.}
\bvolume{951}(\bissue{1}),
\bfpage{8}
(\byear{2023})
\doiurl{10.3847/2041-8213/acdac6}
{\href{https://arxiv.org/abs/2306.16213}{{arXiv:2306.16213}}}
{[astro-ph.HE]}
\end{barticle}
\endbibitem

\bibitem[\protect\citeauthoryear{Antoniadis et~al.}{2023a}]{EPTA:2023sfo}
\begin{barticle}
\bauthor{\bsnm{Antoniadis}, \binits{J.}}, \betal:
\batitle{{The second data release from the European Pulsar Timing Array I. The
  dataset and timing analysis}}.
\bjtitle{Astron. Astrophys.}
\bvolume{678},
\bfpage{48}
(\byear{2023})
\doiurl{10.1051/0004-6361/202346841}
{\href{https://arxiv.org/abs/2306.16224}{{arXiv:2306.16224}}}
{[astro-ph.HE]}
\end{barticle}
\endbibitem

\bibitem[\protect\citeauthoryear{Antoniadis et~al.}{2023b}]{EPTA:2023akd}
\begin{barticle}
\bauthor{\bsnm{Antoniadis}, \binits{J.}}, \betal:
\batitle{{The second data release from the European Pulsar Timing Array II.
  Customised pulsar noise models for spatially correlated gravitational
  waves}}.
\bjtitle{Astron. Astrophys.}
\bvolume{678},
\bfpage{49}
(\byear{2023})
\doiurl{10.1051/0004-6361/202346842}
{\href{https://arxiv.org/abs/2306.16225}{{arXiv:2306.16225}}}
{[astro-ph.HE]}
\end{barticle}
\endbibitem

\bibitem[\protect\citeauthoryear{Antoniadis et~al.}{2023c}]{EPTA:2023fyk}
\begin{barticle}
\bauthor{\bsnm{Antoniadis}, \binits{J.}}, \betal:
\batitle{{The second data release from the European Pulsar Timing Array III.
  Search for gravitational wave signals}}.
\bjtitle{Astron. Astrophys.}
\bvolume{678},
\bfpage{50}
(\byear{2023})
\doiurl{10.1051/0004-6361/202346844}
{\href{https://arxiv.org/abs/2306.16214}{{arXiv:2306.16214}}}
{[astro-ph.HE]}
\end{barticle}
\endbibitem

\bibitem[\protect\citeauthoryear{Zic et~al.}{2023}]{Zic:2023gta}
\begin{barticle}
\bauthor{\bsnm{Zic}, \binits{A.}}, \betal:
\batitle{{The Parkes Pulsar Timing Array third data release}}.
\bjtitle{Publ. Astron. Soc. Austral.}
\bvolume{40},
\bfpage{049}
(\byear{2023})
\doiurl{10.1017/pasa.2023.36}
{\href{https://arxiv.org/abs/2306.16230}{{arXiv:2306.16230}}}
{[astro-ph.HE]}
\end{barticle}
\endbibitem

\bibitem[\protect\citeauthoryear{Reardon et~al.}{2023}]{Reardon:2023gzh}
\begin{barticle}
\bauthor{\bsnm{Reardon}, \binits{D.J.}}, \betal:
\batitle{{Search for an Isotropic Gravitational-wave Background with the Parkes
  Pulsar Timing Array}}.
\bjtitle{Astrophys. J. Lett.}
\bvolume{951}(\bissue{1}),
\bfpage{6}
(\byear{2023})
\doiurl{10.3847/2041-8213/acdd02}
{\href{https://arxiv.org/abs/2306.16215}{{arXiv:2306.16215}}}
{[astro-ph.HE]}
\end{barticle}
\endbibitem

\bibitem[\protect\citeauthoryear{Xu et~al.}{2023}]{Xu:2023wog}
\begin{barticle}
\bauthor{\bsnm{Xu}, \binits{H.}}, \betal:
\batitle{{Searching for the Nano-Hertz Stochastic Gravitational Wave Background
  with the Chinese Pulsar Timing Array Data Release I}}.
\bjtitle{Res. Astron. Astrophys.}
\bvolume{23}(\bissue{7}),
\bfpage{075024}
(\byear{2023})
\doiurl{10.1088/1674-4527/acdfa5}
{\href{https://arxiv.org/abs/2306.16216}{{arXiv:2306.16216}}}
{[astro-ph.HE]}
\end{barticle}
\endbibitem

\bibitem[\protect\citeauthoryear{Punturo et~al.}{2010}]{Punturo:2010zz}
\begin{barticle}
\bauthor{\bsnm{Punturo}, \binits{M.}}, \betal:
\batitle{{The Einstein Telescope: A third-generation gravitational wave
  observatory}}.
\bjtitle{Class. Quant. Grav.}
\bvolume{27},
\bfpage{194002}
(\byear{2010})
\doiurl{10.1088/0264-9381/27/19/194002}
\end{barticle}
\endbibitem

\bibitem[\protect\citeauthoryear{Reitze et~al.}{2019}]{Reitze:2019iox}
\begin{barticle}
\bauthor{\bsnm{Reitze}, \binits{D.}}, \betal:
\batitle{{Cosmic Explorer: The U.S. Contribution to Gravitational-Wave
  Astronomy beyond LIGO}}.
\bjtitle{Bull. Am. Astron. Soc.}
\bvolume{51}(\bissue{7}),
\bfpage{035}
(\byear{2019})
{\href{https://arxiv.org/abs/1907.04833}{{arXiv:1907.04833}}}
{[astro-ph.IM]}
\end{barticle}
\endbibitem

\bibitem[\protect\citeauthoryear{Amaro-Seoane et~al.}{2017}]{LISA:2017pwj}
\begin{botherref}
\oauthor{\bsnm{Amaro-Seoane}, \binits{P.}}, et al.:
{Laser Interferometer Space Antenna}.
e-prints
(2017)
{\href{https://arxiv.org/abs/1702.00786}{{arXiv:1702.00786}}}
{[astro-ph.IM]}
\end{botherref}
\endbibitem

\bibitem[\protect\citeauthoryear{Hu and Wu}{2017}]{Hu:2017mde}
\begin{barticle}
\bauthor{\bsnm{Hu}, \binits{W.-R.}},
\bauthor{\bsnm{Wu}, \binits{Y.-L.}}:
\batitle{{The Taiji Program in Space for gravitational wave physics and the
  nature of gravity}}.
\bjtitle{Natl. Sci. Rev.}
\bvolume{4}(\bissue{5}),
\bfpage{685}--\blpage{686}
(\byear{2017})
\doiurl{10.1093/nsr/nwx116}
\end{barticle}
\endbibitem

\bibitem[\protect\citeauthoryear{Luo et~al.}{2016}]{TianQin:2015yph}
\begin{barticle}
\bauthor{\bsnm{Luo}, \binits{J.}}, \betal:
\batitle{{TianQin: a space-borne gravitational wave detector}}.
\bjtitle{Class. Quant. Grav.}
\bvolume{33}(\bissue{3}),
\bfpage{035010}
(\byear{2016})
\doiurl{10.1088/0264-9381/33/3/035010}
{\href{https://arxiv.org/abs/1512.02076}{{arXiv:1512.02076}}}
{[astro-ph.IM]}
\end{barticle}
\endbibitem

\bibitem[\protect\citeauthoryear{Kawamura et~al.}{2011}]{Kawamura:2011zz}
\begin{barticle}
\bauthor{\bsnm{Kawamura}, \binits{S.}}, \betal:
\batitle{{The Japanese space gravitational wave antenna: DECIGO}}.
\bjtitle{Class. Quant. Grav.}
\bvolume{28},
\bfpage{094011}
(\byear{2011})
\doiurl{10.1088/0264-9381/28/9/094011}
\end{barticle}
\endbibitem

\bibitem[\protect\citeauthoryear{Harms et~al.}{2021}]{LGWA:2020mma}
\begin{barticle}
\bauthor{\bsnm{Harms}, \binits{J.}}, \betal:
\batitle{{Lunar Gravitational-wave Antenna}}.
\bjtitle{Astrophys. J.}
\bvolume{910}(\bissue{1}),
\bfpage{1}
(\byear{2021})
\doiurl{10.3847/1538-4357/abe5a7}
{\href{https://arxiv.org/abs/2010.13726}{{arXiv:2010.13726}}}
{[gr-qc]}
\end{barticle}
\endbibitem

\bibitem[\protect\citeauthoryear{Ajith et~al.}{2025}]{Ajith:2024mie}
\begin{barticle}
\bauthor{\bsnm{Ajith}, \binits{P.}}, \betal:
\batitle{{The Lunar Gravitational-wave Antenna: mission studies and science
  case}}.
\bjtitle{JCAP}
\bvolume{01},
\bfpage{108}
(\byear{2025})
\doiurl{10.1088/1475-7516/2025/01/108}
{\href{https://arxiv.org/abs/2404.09181}{{arXiv:2404.09181}}}
{[gr-qc]}
\end{barticle}
\endbibitem

\bibitem[\protect\citeauthoryear{Li et~al.}{2023}]{Li:2023plm}
\begin{barticle}
\bauthor{\bsnm{Li}, \binits{J.}},
\bauthor{\bsnm{Liu}, \binits{F.}},
\bauthor{\bsnm{Pan}, \binits{Y.}},
\bauthor{\bsnm{Wang}, \binits{Z.}},
\bauthor{\bsnm{Cao}, \binits{M.}},
\bauthor{\bsnm{Wang}, \binits{M.}},
\bauthor{\bsnm{Zhang}, \binits{F.}},
\bauthor{\bsnm{Zhang}, \binits{J.}},
\bauthor{\bsnm{Zhu}, \binits{Z.-H.}}:
\batitle{{Detecting gravitational wave with an interferometric seismometer
  array on lunar nearside}}.
\bjtitle{Sci. China Phys. Mech. Astron.}
\bvolume{66}(\bissue{10}),
\bfpage{109513}
(\byear{2023})
\doiurl{10.1007/s11433-023-2179-9} .
\bcomment{[Erratum: Sci.China Phys.Mech.Astron. 67, 219551 (2024)]}
\end{barticle}
\endbibitem

\bibitem[\protect\citeauthoryear{Greene et~al.}{2020}]{Greene:2019vlv}
\begin{barticle}
\bauthor{\bsnm{Greene}, \binits{J.E.}},
\bauthor{\bsnm{Strader}, \binits{J.}},
\bauthor{\bsnm{Ho}, \binits{L.C.}}:
\batitle{{Intermediate-Mass Black Holes}}.
\bjtitle{Ann. Rev. Astron. Astrophys.}
\bvolume{58},
\bfpage{257}--\blpage{312}
(\byear{2020})
\doiurl{10.1146/annurev-astro-032620-021835}
{\href{https://arxiv.org/abs/1911.09678}{{arXiv:1911.09678}}}
{[astro-ph.GA]}
\end{barticle}
\endbibitem

\bibitem[\protect\citeauthoryear{Mezcua}{2017}]{Mezcua:2017npy}
\begin{barticle}
\bauthor{\bsnm{Mezcua}, \binits{M.}}:
\batitle{{Observational evidence for intermediate-mass black holes}}.
\bjtitle{Int. J. Mod. Phys. D}
\bvolume{26}(\bissue{11}),
\bfpage{1730021}
(\byear{2017})
\doiurl{10.1142/S021827181730021X}
{\href{https://arxiv.org/abs/1705.09667}{{arXiv:1705.09667}}}
{[astro-ph.GA]}
\end{barticle}
\endbibitem

\bibitem[\protect\citeauthoryear{Maggiore}{2007}]{Maggiore:2007ulw}
\begin{bbook}
\bauthor{\bsnm{Maggiore}, \binits{M.}}:
\bbtitle{Gravitational Waves. Vol. 1: Theory and Experiments}.
\bpublisher{Oxford University Press},
\blocation{Oxford}
(\byear{2007}).
\doiurl{10.1093/acprof:oso/9780198570745.001.0001}
\end{bbook}
\endbibitem

\bibitem[\protect\citeauthoryear{Will}{2004}]{Will:2004fj}
\begin{barticle}
\bauthor{\bsnm{Will}, \binits{C.M.}}:
\batitle{{On the rate of detectability of intermediate-mass black-hole binaries
  using LISA}}.
\bjtitle{Astrophys. J.}
\bvolume{611},
\bfpage{1080}
(\byear{2004})
\doiurl{10.1086/422387}
{\href{https://arxiv.org/abs/astro-ph/0403644}{{arXiv:astro-ph/0403644}}}
\end{barticle}
\endbibitem

\bibitem[\protect\citeauthoryear{Arca-Sedda et~al.}{2021}]{Arca-Sedda:2020lso}
\begin{barticle}
\bauthor{\bsnm{Arca-Sedda}, \binits{M.}},
\bauthor{\bsnm{Amaro-Seoane}, \binits{P.}},
\bauthor{\bsnm{Chen}, \binits{X.}}:
\batitle{{Merging stellar and intermediate-mass black holes in dense clusters:
  implications for LIGO, LISA, and the next generation of gravitational wave
  detectors}}.
\bjtitle{Astron. Astrophys.}
\bvolume{652},
\bfpage{54}
(\byear{2021})
\doiurl{10.1051/0004-6361/202037785}
{\href{https://arxiv.org/abs/2007.13746}{{arXiv:2007.13746}}}
{[astro-ph.GA]}
\end{barticle}
\endbibitem

\bibitem[\protect\citeauthoryear{Strokov et~al.}{2023}]{Strokov:2023kmo}
\begin{barticle}
\bauthor{\bsnm{Strokov}, \binits{V.}},
\bauthor{\bsnm{Fragione}, \binits{G.}},
\bauthor{\bsnm{Berti}, \binits{E.}}:
\batitle{{LISA constraints on an intermediate-mass black hole in the Galactic
  Centre}}.
\bjtitle{Mon. Not. Roy. Astron. Soc.}
\bvolume{524}(\bissue{2}),
\bfpage{2033}--\blpage{2041}
(\byear{2023})
\doiurl{10.1093/mnras/stad2002}
{\href{https://arxiv.org/abs/2303.00015}{{arXiv:2303.00015}}}
{[astro-ph.HE]}
\end{barticle}
\endbibitem

\bibitem[\protect\citeauthoryear{Liu et~al.}{2024}]{Liu:2023zea}
\begin{barticle}
\bauthor{\bsnm{Liu}, \binits{S.}},
\bauthor{\bsnm{Wang}, \binits{L.}},
\bauthor{\bsnm{Hu}, \binits{Y.-M.}},
\bauthor{\bsnm{Tanikawa}, \binits{A.}},
\bauthor{\bsnm{Trani}, \binits{A.A.}}:
\batitle{{Merging hierarchical triple black hole systems with intermediate-mass
  black holes in population III star clusters}}.
\bjtitle{Mon. Not. Roy. Astron. Soc.}
\bvolume{533}(\bissue{2}),
\bfpage{2262}--\blpage{2281}
(\byear{2024})
\doiurl{10.1093/mnras/stae1946}
{\href{https://arxiv.org/abs/2311.05393}{{arXiv:2311.05393}}}
{[astro-ph.GA]}
\end{barticle}
\endbibitem

\bibitem[\protect\citeauthoryear{Sedda et~al.}{2020}]{Sedda:2019uro}
\begin{barticle}
\bauthor{\bsnm{Sedda}, \binits{M.A.}}, \betal:
\batitle{{The missing link in gravitational-wave astronomy: discoveries waiting
  in the decihertz range}}.
\bjtitle{Class. Quant. Grav.}
\bvolume{37}(\bissue{21}),
\bfpage{215011}
(\byear{2020})
\doiurl{10.1088/1361-6382/abb5c1}
{\href{https://arxiv.org/abs/1908.11375}{{arXiv:1908.11375}}}
{[gr-qc]}
\end{barticle}
\endbibitem

\bibitem[\protect\citeauthoryear{Graff et~al.}{2015}]{Graff:2015bba}
\begin{barticle}
\bauthor{\bsnm{Graff}, \binits{P.B.}},
\bauthor{\bsnm{Buonanno}, \binits{A.}},
\bauthor{\bsnm{Sathyaprakash}, \binits{B.S.}}:
\batitle{{Missing Link: Bayesian detection and measurement of intermediate-mass
  black-hole binaries}}.
\bjtitle{Phys. Rev. D}
\bvolume{92}(\bissue{2}),
\bfpage{022002}
(\byear{2015})
\doiurl{10.1103/PhysRevD.92.022002}
{\href{https://arxiv.org/abs/1504.04766}{{arXiv:1504.04766}}}
{[gr-qc]}
\end{barticle}
\endbibitem

\bibitem[\protect\citeauthoryear{Veitch et~al.}{2015}]{Veitch:2015ela}
\begin{barticle}
\bauthor{\bsnm{Veitch}, \binits{J.}},
\bauthor{\bsnm{P\"urrer}, \binits{M.}},
\bauthor{\bsnm{Mandel}, \binits{I.}}:
\batitle{{Measuring intermediate mass black hole binaries with advanced
  gravitational wave detectors}}.
\bjtitle{Phys. Rev. Lett.}
\bvolume{115}(\bissue{14}),
\bfpage{141101}
(\byear{2015})
\doiurl{10.1103/PhysRevLett.115.141101}
{\href{https://arxiv.org/abs/1503.05953}{{arXiv:1503.05953}}}
{[astro-ph.HE]}
\end{barticle}
\endbibitem

\bibitem[\protect\citeauthoryear{Han et~al.}{2017}]{Han:2017evx}
\begin{barticle}
\bauthor{\bsnm{Han}, \binits{W.-B.}},
\bauthor{\bsnm{Cao}, \binits{Z.}},
\bauthor{\bsnm{Hu}, \binits{Y.-M.}}:
\batitle{{Excitation of high frequency voices from intermediate-mass-ratio
  inspirals with large eccentricity}}.
\bjtitle{Class. Quant. Grav.}
\bvolume{34}(\bissue{22}),
\bfpage{225010}
(\byear{2017})
\doiurl{10.1088/1361-6382/aa891b}
{\href{https://arxiv.org/abs/1710.00147}{{arXiv:1710.00147}}}
{[gr-qc]}
\end{barticle}
\endbibitem

\bibitem[\protect\citeauthoryear{Huerta and Gair}{2011}]{Huerta:2010tp}
\begin{barticle}
\bauthor{\bsnm{Huerta}, \binits{E.A.}},
\bauthor{\bsnm{Gair}, \binits{J.R.}}:
\batitle{{Intermediate-mass-ratio-inspirals in the Einstein Telescope. II.
  Parameter estimation errors}}.
\bjtitle{Phys. Rev. D}
\bvolume{83},
\bfpage{044021}
(\byear{2011})
\doiurl{10.1103/PhysRevD.83.044021}
{\href{https://arxiv.org/abs/1011.0421}{{arXiv:1011.0421}}}
{[gr-qc]}
\end{barticle}
\endbibitem

\bibitem[\protect\citeauthoryear{Reali et~al.}{2024}]{Reali:2024hqf}
\begin{barticle}
\bauthor{\bsnm{Reali}, \binits{L.}},
\bauthor{\bsnm{Cotesta}, \binits{R.}},
\bauthor{\bsnm{Antonelli}, \binits{A.}},
\bauthor{\bsnm{Kritos}, \binits{K.}},
\bauthor{\bsnm{Strokov}, \binits{V.}},
\bauthor{\bsnm{Berti}, \binits{E.}}:
\batitle{{Intermediate-mass black hole binary parameter estimation with
  next-generation ground-based detector networks}}.
\bjtitle{Phys. Rev. D}
\bvolume{110}(\bissue{10}),
\bfpage{103002}
(\byear{2024})
\doiurl{10.1103/PhysRevD.110.103002}
{\href{https://arxiv.org/abs/2406.01687}{{arXiv:2406.01687}}}
{[gr-qc]}
\end{barticle}
\endbibitem

\bibitem[\protect\citeauthoryear{Garc\'\i{}a-Quir\'os
  et~al.}{2020}]{Garcia-Quiros:2020qpx}
\begin{barticle}
\bauthor{\bsnm{Garc\'\i{}a-Quir\'os}, \binits{C.}},
\bauthor{\bsnm{Colleoni}, \binits{M.}},
\bauthor{\bsnm{Husa}, \binits{S.}},
\bauthor{\bsnm{Estell\'es}, \binits{H.}},
\bauthor{\bsnm{Pratten}, \binits{G.}},
\bauthor{\bsnm{Ramos-Buades}, \binits{A.}},
\bauthor{\bsnm{Mateu-Lucena}, \binits{M.}},
\bauthor{\bsnm{Jaume}, \binits{R.}}:
\batitle{{Multimode frequency-domain model for the gravitational wave signal
  from nonprecessing black-hole binaries}}.
\bjtitle{Phys. Rev. D}
\bvolume{102}(\bissue{6}),
\bfpage{064002}
(\byear{2020})
\doiurl{10.1103/PhysRevD.102.064002}
{\href{https://arxiv.org/abs/2001.10914}{{arXiv:2001.10914}}}
{[gr-qc]}
\end{barticle}
\endbibitem

\bibitem[\protect\citeauthoryear{Whelan}{2013}]{whelan2013geometry}
\begin{botherref}
\oauthor{\bsnm{Whelan}, \binits{J.T.}}:
The geometry of gravitational wave detection.
\href{https://dcc-llo.ligo.org/public/0106/T1300666/003/Whelan_geometry.pdf}{LIGO
  Document Control Center}
(2013)
\end{botherref}
\endbibitem

\bibitem[\protect\citeauthoryear{Branchesi et~al.}{2023}]{Branchesi:2023sjl}
\begin{barticle}
\bauthor{\bsnm{Branchesi}, \binits{M.}}, \betal:
\batitle{{Lunar Gravitational-Wave Detection}}.
\bjtitle{Space Sci. Rev.}
\bvolume{219}(\bissue{8}),
\bfpage{67}
(\byear{2023})
\doiurl{10.1007/s11214-023-01015-4}
\end{barticle}
\endbibitem

\bibitem[\protect\citeauthoryear{Dupletsa et~al.}{2023}]{Dupletsa:2022scg}
\begin{barticle}
\bauthor{\bsnm{Dupletsa}, \binits{U.}},
\bauthor{\bsnm{Harms}, \binits{J.}},
\bauthor{\bsnm{Banerjee}, \binits{B.}},
\bauthor{\bsnm{Branchesi}, \binits{M.}},
\bauthor{\bsnm{Goncharov}, \binits{B.}},
\bauthor{\bsnm{Maselli}, \binits{A.}},
\bauthor{\bsnm{Oliveira}, \binits{A.C.S.}},
\bauthor{\bsnm{Ronchini}, \binits{S.}},
\bauthor{\bsnm{Tissino}, \binits{J.}}:
\batitle{{gwfish: A simulation software to evaluate parameter-estimation
  capabilities of gravitational-wave detector networks}}.
\bjtitle{Astron. Comput.}
\bvolume{42},
\bfpage{100671}
(\byear{2023})
\doiurl{10.1016/j.ascom.2022.100671}
{\href{https://arxiv.org/abs/2205.02499}{{arXiv:2205.02499}}}
{[gr-qc]}
\end{barticle}
\endbibitem

\bibitem[\protect\citeauthoryear{{Yan} et~al.}{2024}]{Yan:2024sgwb}
\begin{barticle}
\bauthor{\bsnm{{Yan}}, \binits{H.}},
\bauthor{\bsnm{{Chen}}, \binits{X.}},
\bauthor{\bsnm{{Zhang}}, \binits{J.}},
\bauthor{\bsnm{{Zhang}}, \binits{F.}},
\bauthor{\bsnm{{Shao}}, \binits{L.}},
\bauthor{\bsnm{{Wang}}, \binits{M.}}:
\batitle{{Constraining the stochastic gravitational wave background using the
  future lunar seismometers}}.
\bjtitle{Phys. Rev. D}
\bvolume{110}(\bissue{4}),
\bfpage{043009}
(\byear{2024})
\doiurl{10.1103/PhysRevD.110.043009}
{\href{https://arxiv.org/abs/2405.12640}{{arXiv:2405.12640}}}
{[gr-qc]}
\end{barticle}
\endbibitem

\bibitem[\protect\citeauthoryear{Finn}{1992}]{Finn:1992wt}
\begin{barticle}
\bauthor{\bsnm{Finn}, \binits{L.S.}}:
\batitle{{Detection, measurement and gravitational radiation}}.
\bjtitle{Phys. Rev. D}
\bvolume{46},
\bfpage{5236}--\blpage{5249}
(\byear{1992})
\doiurl{10.1103/PhysRevD.46.5236}
{\href{https://arxiv.org/abs/gr-qc/9209010}{{arXiv:gr-qc/9209010}}}
\end{barticle}
\endbibitem

\bibitem[\protect\citeauthoryear{Borhanian}{2021}]{Borhanian:2020ypi}
\begin{barticle}
\bauthor{\bsnm{Borhanian}, \binits{S.}}:
\batitle{{GWBENCH: a novel Fisher information package for gravitational-wave
  benchmarking}}.
\bjtitle{Class. Quant. Grav.}
\bvolume{38}(\bissue{17}),
\bfpage{175014}
(\byear{2021})
\doiurl{10.1088/1361-6382/ac1618}
{\href{https://arxiv.org/abs/2010.15202}{{arXiv:2010.15202}}}
{[gr-qc]}
\end{barticle}
\endbibitem

\bibitem[\protect\citeauthoryear{Moore et~al.}{2015}]{Moore:2014lga}
\begin{barticle}
\bauthor{\bsnm{Moore}, \binits{C.J.}},
\bauthor{\bsnm{Cole}, \binits{R.H.}},
\bauthor{\bsnm{Berry}, \binits{C.P.L.}}:
\batitle{{Gravitational-wave sensitivity curves}}.
\bjtitle{Class. Quant. Grav.}
\bvolume{32}(\bissue{1}),
\bfpage{015014}
(\byear{2015})
\doiurl{10.1088/0264-9381/32/1/015014}
{\href{https://arxiv.org/abs/1408.0740}{{arXiv:1408.0740}}}
{[gr-qc]}
\end{barticle}
\endbibitem

\bibitem[\protect\citeauthoryear{Li et~al.}{2022}]{Li:2021mbo}
\begin{barticle}
\bauthor{\bsnm{Li}, \binits{Y.}},
\bauthor{\bsnm{Heng}, \binits{I.S.}},
\bauthor{\bsnm{Chan}, \binits{M.L.}},
\bauthor{\bsnm{Messenger}, \binits{C.}},
\bauthor{\bsnm{Fan}, \binits{X.}}:
\batitle{{Exploring the sky localization and early warning capabilities of
  third generation gravitational wave detectors in three-detector network
  configurations}}.
\bjtitle{Phys. Rev. D}
\bvolume{105}(\bissue{4}),
\bfpage{043010}
(\byear{2022})
\doiurl{10.1103/PhysRevD.105.043010}
{\href{https://arxiv.org/abs/2109.07389}{{arXiv:2109.07389}}}
{[astro-ph.IM]}
\end{barticle}
\endbibitem

\bibitem[\protect\citeauthoryear{Sesana et~al.}{2011}]{Sesana:2010wy}
\begin{barticle}
\bauthor{\bsnm{Sesana}, \binits{A.}},
\bauthor{\bsnm{Gair}, \binits{J.}},
\bauthor{\bsnm{Berti}, \binits{E.}},
\bauthor{\bsnm{Volonteri}, \binits{M.}}:
\batitle{{Reconstructing the massive black hole cosmic history through
  gravitational waves}}.
\bjtitle{Phys. Rev. D}
\bvolume{83},
\bfpage{044036}
(\byear{2011})
\doiurl{10.1103/PhysRevD.83.044036}
{\href{https://arxiv.org/abs/1011.5893}{{arXiv:1011.5893}}}
{[astro-ph.CO]}
\end{barticle}
\endbibitem

\bibitem[\protect\citeauthoryear{Sesana et~al.}{2008}]{Sesana:2007vr}
\begin{barticle}
\bauthor{\bsnm{Sesana}, \binits{A.}},
\bauthor{\bsnm{Haardt}, \binits{F.}},
\bauthor{\bsnm{Madau}, \binits{P.}}:
\batitle{{Interaction of massive black hole binaries with their stellar
  environment. 3. Scattering of bound stars}}.
\bjtitle{Astrophys. J.}
\bvolume{686},
\bfpage{432}
(\byear{2008})
\doiurl{10.1086/590651}
{\href{https://arxiv.org/abs/0710.4301}{{arXiv:0710.4301}}}
{[astro-ph]}
\end{barticle}
\endbibitem

\bibitem[\protect\citeauthoryear{Klein et~al.}{2016}]{Klein:2015hvg}
\begin{barticle}
\bauthor{\bsnm{Klein}, \binits{A.}}, \betal:
\batitle{{Science with the space-based interferometer eLISA: Supermassive black
  hole binaries}}.
\bjtitle{Phys. Rev. D}
\bvolume{93}(\bissue{2}),
\bfpage{024003}
(\byear{2016})
\doiurl{10.1103/PhysRevD.93.024003}
{\href{https://arxiv.org/abs/1511.05581}{{arXiv:1511.05581}}}
{[gr-qc]}
\end{barticle}
\endbibitem

\bibitem[\protect\citeauthoryear{Yan et~al.}{2024}]{Yan:2024jio}
\begin{barticle}
\bauthor{\bsnm{Yan}, \binits{H.}},
\bauthor{\bsnm{Chen}, \binits{X.}},
\bauthor{\bsnm{Zhang}, \binits{J.}},
\bauthor{\bsnm{Zhang}, \binits{F.}},
\bauthor{\bsnm{Wang}, \binits{M.}},
\bauthor{\bsnm{Shao}, \binits{L.}}:
\batitle{{Toward a consistent calculation of the lunar response to
  gravitational waves}}.
\bjtitle{Phys. Rev. D}
\bvolume{109}(\bissue{6}),
\bfpage{064092}
(\byear{2024})
\doiurl{10.1103/PhysRevD.109.064092}
{\href{https://arxiv.org/abs/2403.08681}{{arXiv:2403.08681}}}
{[gr-qc]}.
\bcomment{[Erratum: Phys.Rev.D 109, 089903 (2024)]}
\end{barticle}
\endbibitem

\bibitem[\protect\citeauthoryear{{Belgacem} et~al.}{2024}]{Belgacem:2024eft}
\begin{barticle}
\bauthor{\bsnm{{Belgacem}}, \binits{E.}},
\bauthor{\bsnm{{Maggiore}}, \binits{M.}},
\bauthor{\bsnm{{Moreau}}, \binits{T.}}:
\batitle{{Coupling elastic media to gravitational waves: an effective field
  theory approach}}.
\bjtitle{Journal of Cosmology and Astroparticle Physics}
\bvolume{2024}(\bissue{7}),
\bfpage{028}
(\byear{2024})
\doiurl{10.1088/1475-7516/2024/07/028}
{\href{https://arxiv.org/abs/2403.16550}{{arXiv:2403.16550}}}
{[gr-qc]}
\end{barticle}
\endbibitem

\bibitem[\protect\citeauthoryear{{Majstorovi{\'c}}
  et~al.}{2024}]{Majstorovic:2024arXiv}
\begin{botherref}
\oauthor{\bsnm{{Majstorovi{\'c}}}, \binits{J.}},
\oauthor{\bsnm{{Vidal}}, \binits{L.}},
\oauthor{\bsnm{{Lognonn{\'e}}}, \binits{P.}}:
{Modeling lunar response to gravitational waves using normal-mode approach and
  tidal forcing}.
arXiv e-prints,
2411--09559
(2024)
\doiurl{10.48550/arXiv.2411.09559}
{\href{https://arxiv.org/abs/2411.09559}{{arXiv:2411.09559}}}
{[gr-qc]}
\end{botherref}
\endbibitem

\end{thebibliography}

\end{document}